\documentclass[aps,prb,reprint,amsmath,amssymb,superscriptaddress,nofootinbib]{revtex4-2}
\usepackage{graphicx}
\usepackage{bm}
\usepackage{hyperref}
\usepackage{xcolor}
\usepackage{physics}
\usepackage{mathtools}

\hypersetup{colorlinks = true, 
	linkcolor = blue, 
	urlcolor = blue,
	citecolor = blue} 
	
\newcommand{\ii}{\mathrm{i}}
\newcommand{\e}{\mathrm{e}}
\newcommand{\I}{\mathbb{I}}
\newcommand{\UT}{\hat U_T}
\newcommand{\Wone}{\hat W_1}
\newcommand{\Wtwo}{\hat W_2}
\newcommand{\ketA}{\ket{A}}
\newcommand{\ketB}{\ket{B}}
\newcommand{\Cbar}{\overline{C}}

\begin{document}

\title{Boundary $0/\pi$ logical subspace and bulk dynamical probes in flux-controlled anomalous Floquet quantum walks}

\author{WeiCheng Ning}
\affiliation{College of Physics and Materials Science, Tianjin Normal University, Tianjin 300387, China}

\author{YanSheng Liu}
\affiliation{College of Physics and Materials Science, Tianjin Normal University, Tianjin 300387, China}

\author{XiaoXue Zhang}
\affiliation{College of Physics and Materials Science, Tianjin Normal University, Tianjin 300387, China}

\author{XiZheng Zhang}
\email{zhangxz@tjnu.edu.cn}
\affiliation{College of Physics and Materials Science, Tianjin Normal University, Tianjin 300387, China}


\begin{abstract}
We formulate a one-dimensional flux-controlled anomalous Floquet quantum walk and show that it admits a direct microscopic realization in a driven bipartite lattice. The walk consists of a coin-dependent drift step and a momentum-dependent coin mixing step, so the same evolution operator governs quasienergy bands, boundary modes, and bulk dynamics in real space. Because the walk is chiral, the quasienergy gaps at $0$ and $\pi/T$ carry independent topological information, which organizes trivial, $0$-only, $\pi$-only, and coexistence sectors in the $(M,\phi)$ plane. In the coexistence sector, a $0$ mode and a $\pi$ mode reside on the same edge and span a natural boundary logical subspace. One Floquet period acts there as a relative phase operation and produces a clear $2T$ response in local boundary observables. In the bulk, the same anomalous Floquet structure is probed dynamically in two complementary ways. Frame-resolved mean chiral displacements approach the two winding numbers in the clean pre-reflection window of the symmetric time frames, while selected benchmark cuts at a representative $0$ gap closing and a representative $\pi$ gap closing exhibit distinct local stroboscopic responses, with the $\pi$ gap benchmark showing a much stronger odd-even alternation. The boundary logical subspace and the bulk dynamical probes are therefore organized within one flux-controlled anomalous Floquet quantum walk, suggesting a symmetry-protected route to quantum-walk information primitives in driven microstructured lattices.
\end{abstract}

\maketitle

\section{Introduction}

Discrete-time quantum walks provide a natural framework for periodically driven topology because a single step operator simultaneously determines the quasienergy bands, the boundary spectrum, and the stroboscopic dynamics~\cite{eckardtHighfrequencyApproximationPeriodically2015,harperTopologyBrokenSymmetry2020,kitagawaTopologicalCharacterizationPeriodically2010,okaFloquetEngineeringQuantum2019,rudnerBandStructureEngineering2020,zhouNonHermitianFloquetTopological2018,asbothChiralSymmetryBulkboundary2014,cedzichChiralFloquetSystems2021,xuExperimentalClassificationQuenched2019}. In one dimension, chiral Floquet walks are distinguished by the fact that the quasienergy gaps at $0$ and $\pi/T$ can carry independent topological information~\cite{naDistinctFloquetTopological,rudnerAnomalousEdgeStates2013,zhouNonHermitianFloquetTopological2018,asbothChiralSymmetryBulkboundary2014}. This feature is encoded by winding numbers defined in two symmetric time frames, whose combinations determine the numbers of edge modes pinned at quasienergies $0$ and $\pi/T$~\cite{kitagawaExploringTopologicalPhases2010,kitagawaTopologicalPhenomenaQuantum2012,asbothBulkBoundaryCorrespondenceChiral2013,asbothSymmetriesTopologicalPhases2012,bomantaraQuantumComputationFloquet2018,x9kx-v9d2}.

The present work is organized from that quantum walk viewpoint from the outset. We identify a flux-controlled Floquet walk whose drift step is coin-dependent and whose mixing step is momentum-dependent, and then show that this walk is realized exactly by a driven bipartite lattice~\cite{kitagawaTopologicalCharacterizationPeriodically2010}. The required ingredients, namely a bipartite lattice, sublattice-resolved hopping, periodic driving, and a controllable Peierls phase, are compatible with existing quantum-walk and Floquet-lattice platforms~\cite{eckardtHighfrequencyApproximationPeriodically2015,jotzuExperimentalRealisationTopological2014,winterspergerRealizationAnomalousFloquet2020,cardanoDetectionZakPhases2017}. This order of presentation is essential for the logic of the paper. Once the evolution is written as a Floquet walk, the natural observables are not limited to quasienergy gaps and open boundary spectra. One should also ask how the dual gap topology is read out dynamically in boundary and bulk measurements~\cite{asbothBulkBoundaryCorrespondenceChiral2013,zhouNonHermitianFloquetTopological2018}.

Within this language the central structure becomes sharp. In the bulk, the chiral Floquet walk requires two winding numbers and therefore two gap indices, one for the $0$ gap and one for the $\pi/T$ gap~\cite{asbothChiralSymmetryBulkboundary2014}. Under open boundary conditions, the coexistence sector hosts a $0$ mode and a $\pi$ mode on the same edge. Their span defines a minimal boundary logical subspace in which one period of Floquet evolution produces a relative sign. That relative sign is directly resolved as a $2T$ response of local edge observables. The bulk sector should then be treated in the same spirit. Rather than emphasizing generic packet steering~\cite{huangNondispersingWavePackets2021,zhaoPhaseModulationDirected2024,winterspergerRealizationAnomalousFloquet2020,vuDynamicBulkboundaryCorrespondence2022}, we focus on dynamical probes that read out the anomalous Floquet topology itself. One probe is frame-resolved mean chiral displacement, which tracks the winding numbers in the two symmetric time frames~\cite{cardanoDetectionZakPhases2017,derricoBulkDetectionTimedependent2020,maffeiTopologicalCharacterizationChiral2018}. The other is the short-to-intermediate-time local response near representative $0$ gap and $\pi$ gap critical points. This distinction is useful because the two gap closings are not dynamically equivalent in stroboscopic evolution~\cite{panUniversalPresenceTimecrystalline,naDistinctFloquetTopological}.

The paper addresses a more specific question than the construction of another driven-lattice phase diagram. We ask whether one microscopic family can support a coherent set of anomalous Floquet quantum-walk structures in which boundary and bulk diagnostics are controlled within the same flux-tuned protocol. In this formulation the coexistence sector furnishes a boundary $0/\pi$ logical subspace together with a doubled-period edge response, while the bulk sector furnishes dynamical probes of the dual gap topology~\cite{panahiyanControllableSimulationTopological2020,jiaHighWindingNumber2021,unalHowDirectlyMeasure2019,vuDynamicBulkboundaryCorrespondence2022,zhouNonHermitianFloquetTopological2018}. This setting is not meant as a claim of topological quantum computation in the strict non-Abelian sense~\cite{kitaevFaulttolerantQuantumComputation2003,nayakNonAbelianAnyonsTopological2008}, but it does provide a minimal and microscopically controlled quantum-walk platform in which boundary logical structure and bulk dynamical readout coexist.

The remainder of the paper follows that logic. Section II introduces the Floquet walk and derives its microscopic lattice realization. Section III establishes the anomalous bulk topology. Section IV analyzes the boundary $0/\pi$ logical subspace and the doubled-period response. Section V turns to bulk dynamical probes and separates frame-resolved topological readout from selected benchmark critical dynamics. Section VI summarizes the physical picture.

\section{Flux-controlled Floquet quantum walk and microscopic realization}

We consider a one-dimensional walk whose Hilbert space factorizes as
\begin{equation}
\mathcal H=\mathcal H_{\mathrm p}\otimes\mathcal H_{\mathrm c},
\qquad
\mathcal H_{\mathrm c}=\mathrm{span}\{\ketA,\ketB\},
\label{eq:Hilbert_factor}
\end{equation}
where $\mathcal H_{\mathrm p}$ labels the lattice position and the two-dimensional space $\mathcal H_{\mathrm c}$ plays the role of the coin. One Floquet period is built from two substeps,
\begin{equation}
\UT=\Wtwo\Wone,
\qquad
\Wone=\e^{-\ii \frac{T}{2}\hat H_1},
\qquad
\Wtwo=\e^{-\ii \frac{T}{2}\hat H_2}.
\label{eq:UFdef}
\end{equation}
In momentum space the corresponding walk operator will take the form
\begin{equation}
U(k)=\e^{-\ii\theta_2(k)\sigma_x}\e^{-\ii\theta_1(k)\sigma_z}.
\label{eq:Ukcompact}
\end{equation}
The factor generated by $\sigma_z$ is diagonal in the coin basis and produces opposite phases for the two coin components, whereas the factor generated by $\sigma_x$ mixes the two components. Equation \eqref{eq:Ukcompact} is already the defining structure of a Floquet quantum walk, with the drift substep producing the coin-dependent phase motion and the mixing substep rotating the coin.

To expose the real space update rule, let
\begin{equation}
\ket{\Psi(m)}=\sum_n\qty[a_n^{(m)}\ket{n,A}+b_n^{(m)}\ket{n,B}],
\label{eq:state_realspace}
\end{equation}
with Bloch spinor
\begin{equation}
\psi_k^{(m)}=
\begin{pmatrix}
 a_k^{(m)}\\
 b_k^{(m)}
\end{pmatrix}.
\label{eq:spinor_def}
\end{equation}
After the drift substep,
\begin{equation}
\psi_k^{(m+\frac12)}=\e^{-\ii\theta_1(k)\sigma_z}\psi_k^{(m)},
\label{eq:first_half_step}
\end{equation}
so that
\begin{equation}
a_k^{(m+\frac12)}=\e^{-\ii\theta_1(k)}a_k^{(m)},
\qquad
b_k^{(m+\frac12)}=\e^{\ii\theta_1(k)}b_k^{(m)}.
\label{eq:first_half_components}
\end{equation}
The two coin components therefore acquire opposite momentum-dependent phases. After the mixing substep,
\begin{equation}
\psi_k^{(m+1)}=\e^{-\ii\theta_2(k)\sigma_x}\psi_k^{(m+\frac12)},
\label{eq:second_half_step}
\end{equation}
or explicitly
\begin{equation}
\begin{pmatrix}
 a_k^{(m+1)}\\
 b_k^{(m+1)}
\end{pmatrix}
=
\begin{pmatrix}
\cos\theta_2(k) & -\ii\sin\theta_2(k)\\
-\ii\sin\theta_2(k) & \cos\theta_2(k)
\end{pmatrix}
\begin{pmatrix}
 a_k^{(m+\frac12)}\\
 b_k^{(m+\frac12)}
\end{pmatrix}.
\label{eq:coin_mixing_explicit}
\end{equation}
This is a quantum walk step with a momentum-dependent coin rotation. In particular, the phase $\phi$ enters through the drift sector and directly controls how the walk biases the two coin components in momentum space.

We now show that this Floquet walk is realized exactly by a driven bipartite lattice with sublattice operators $a_n$ and $b_n$. The two real space Hamiltonians are chosen as
\begin{align}
\hat H_1={}&-\ii t_1\sum_n\qty(\e^{-\ii\phi}a_{n+1}^\dagger a_n-\e^{\ii\phi}a_n^\dagger a_{n+1})
\nonumber
\\
&+\ii t_1\sum_n\qty(\e^{-\ii\phi}b_{n+1}^\dagger b_n-\e^{\ii\phi}b_n^\dagger b_{n+1}),
\label{eq:H1real}
\end{align}
and
\begin{align}
\hat H_2={}&M\sum_n\qty(a_n^\dagger b_n+b_n^\dagger a_n) 
\nonumber
\\
&+t_2\sum_n\qty(a_{n+1}^\dagger b_n+a_n^\dagger b_{n+1}+\mathrm{H.c.}).
\label{eq:H2real}
\end{align}
Under periodic boundary conditions we introduce the Bloch basis
\begin{equation}
\ket{k}=\frac{1}{\sqrt{L}}\sum_n \e^{-\ii kn}\ket{n},
\qquad k\in[-\pi,\pi),
\label{eq:Bloch_basis}
\end{equation}
which reduces the two substep Hamiltonians to
\begin{equation}
	\begin{split}
	&H_1(k)=2t_1\sin(k-\phi)\,\sigma_z,\\
	&H_2(k)=\qty(M+2t_2\cos k)\sigma_x.
	\label{eq:H1H2k}
	\end{split}
\end{equation}

It is convenient to define
\begin{equation}
	\begin{split}
		&\theta_1(k)=Tt_1\sin(k-\phi),\\
		&\theta_2(k)=\frac{T}{2}\qty(M+2t_2\cos k),
		\label{eq:theta12}
	\end{split}
\end{equation}
which reproduces Eq.~\eqref{eq:Ukcompact}. The explicit Fourier-transform reduction from Eqs.~\eqref{eq:H1real} and \eqref{eq:H2real} to Eq.~\eqref{eq:H1H2k} is given in Appendix~\ref{app:walk_mapping}. The lattice model therefore does not merely resemble a quantum walk. It realizes the walk exactly, with a direct microscopic interpretation of the two substeps. The parameter $\phi$ is a flux phase in the lattice language and a drift control parameter in the quantum walk language.

\section{Anomalous bulk topology}

Because $U(k)$ is a $2\times2$ unitary matrix, it may always be written as
\begin{equation}
U(k)=d_0(k)\I-\ii\,\bm d(k)\cdot\bm\sigma,
\qquad
d_0(k)^2+\abs{\bm d(k)}^2=1.
\label{eq:Udecomp}
\end{equation}
Using Eq.~\eqref{eq:Ukcompact}, one finds
\begin{align}
U(k)={}&\cos\theta_2\cos\theta_1\,\I -\ii\Bigl[\sin\theta_2\cos\theta_1\,\sigma_x\nonumber\\
&-\sin\theta_2\sin\theta_1\,\sigma_y
+\cos\theta_2\sin\theta_1\,\sigma_z\Bigr].
\label{eq:Uexpanded}
\end{align}
The quasienergy eigenvalue equation,
\begin{equation}
U(k)\ket{u_\pm(k)}=\e^{-\ii\varepsilon_\pm(k)T}\ket{u_\pm(k)},
\label{eq:eigeneq}
\end{equation}
gives
\begin{equation}
\cos\qty[\varepsilon(k)T]=\cos\theta_1(k)\cos\theta_2(k),
\label{eq:cosdisp}
\end{equation}
and hence
\begin{equation}
\varepsilon_\pm(k)=\pm\frac{1}{T}\arccos\qty[\cos\theta_1(k)\cos\theta_2(k)].
\label{eq:epsbands}
\end{equation}
The gap closes only when
\begin{equation}
\theta_1(k_*)=m\pi,
\qquad
\theta_2(k_*)=n\pi,
\qquad m,n\in\mathbb Z,
\label{eq:theta_critical}
\end{equation}
which yields the explicit conditions
\begin{equation}
Tt_1\sin(k_* - \phi)=m\pi,
\quad
\frac{T}{2}\qty(M+2t_2\cos k_*)=n\pi.
\label{eq:gapclosing_explicit}
\end{equation}
If $m+n$ is even the closure occurs at quasienergy $0$. If $m+n$ is odd the closure occurs at quasienergy $\pi/T$. The trace derivation of the quasienergy bands and the algebra leading to the gap-closing conditions are collected in Appendix~\ref{app:bands_gaps}.

The walk is chiral because
\begin{equation}
\sigma_y U(k)\sigma_y=U^{-1}(k),
\label{eq:chiral_relation}
\end{equation}
so the chiral operator is
\begin{equation}
\Gamma=\sigma_y.
\label{eq:Gamma_def}
\end{equation}
For one-dimensional chiral Floquet walks, one must work in two symmetric time frames~\cite{asbothBulkBoundaryCorrespondenceChiral2013},
\begin{align}
U_1(k)&=\e^{-\ii\frac{\theta_1(k)}{2}\sigma_z}\e^{-\ii\theta_2(k)\sigma_x}\e^{-\ii\frac{\theta_1(k)}{2}\sigma_z},
\label{eq:U1def}\\
U_2(k)&=\e^{-\ii\frac{\theta_2(k)}{2}\sigma_x}\e^{-\ii\theta_1(k)\sigma_z}\e^{-\ii\frac{\theta_2(k)}{2}\sigma_x}.
\label{eq:U2def}
\end{align}
In these frames,
\begin{equation}
U_\ell(k)=n_{0,\ell}(k)\I-\ii\qty[n_{x,\ell}(k)\sigma_x+n_{z,\ell}(k)\sigma_z],
\label{eq:timeframe_nvec}
\end{equation}
where $\ell=1,2$, with
\begin{align}
n_{x,1}(k)&=\sin\theta_2(k), 
\label{eq:n1x}
\\
n_{z,1}(k)&=\cos\theta_2(k)\sin\theta_1(k),
\label{eq:n1z}\\
n_{x,2}(k)&=\cos\theta_1(k)\sin\theta_2(k), 
\label{eq:n2x}
\\
n_{z,2}(k)&=\sin\theta_1(k).
\label{eq:n2z}
\end{align}
The winding numbers are then
\begin{equation}
W_\ell=\frac{1}{2\pi}\int_{-\pi}^{\pi} dk\,\partial_k\arg\qty[n_{x,\ell}(k)+\ii n_{z,\ell}(k)],
\label{eq:winding_def}
\end{equation}
where $\ell=1,2$ and the gap indices are
\begin{equation}
\nu_0=\frac{W_1+W_2}{2},
\qquad
\nu_\pi=\frac{W_1-W_2}{2}.
\label{eq:nu_def}
\end{equation}
Appendix~\ref{app:winding_derivation} gives the explicit multiplication in the two symmetric time frames and shows how the vectors \((n_{x,\ell},n_{z,\ell})\) define the winding numbers.

\begin{figure*}[t]
\centering
\includegraphics[width=\textwidth]{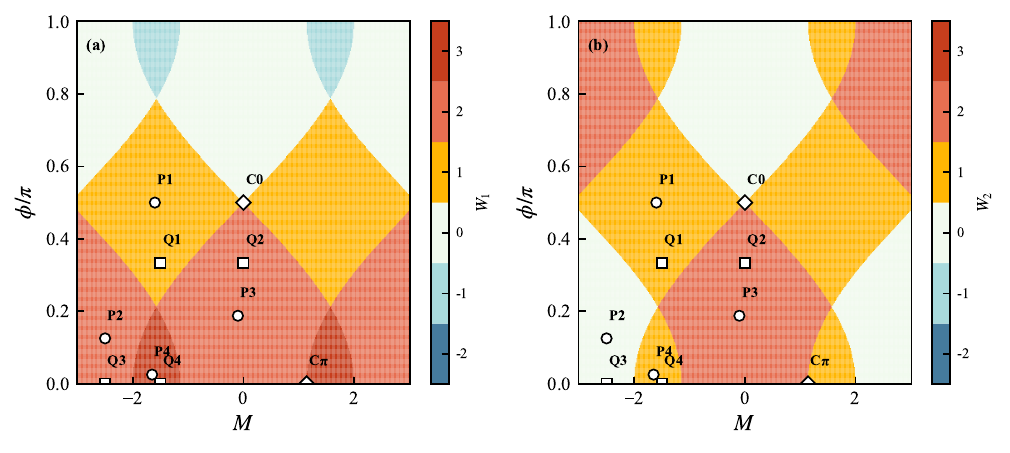}
\caption{Bulk topological phase diagram of the Floquet quantum walk for $t_1=t_2=1$ and $T=2$. Panels (a) and (b) show $W_1$ and $W_2$ evaluated on a high-resolution grid in the $(M,\phi)$ plane. The representative parameter sets used later in Figs.~\ref{fig:edge_dynamics}, \ref{fig:mcd_representative}, and \ref{fig:critical_benchmark} are written directly on the phase map. Circular markers denote the open boundary points $P_j$, square markers denote the frame-resolved bulk points $Q_j$, and diamond markers denote the critical benchmarks $C_0$ and $C_\pi$. Every later numerical panel can therefore be traced back to a definite point in the bulk diagram.}
\label{fig:bulk_phase_map}
\end{figure*}

Figure \ref{fig:bulk_phase_map} shows the resulting bulk structure in the $(M,\phi)$ plane for $t_1=t_2=1$ and $T=2$. The four circularly labeled points are used for the open-boundary results in Fig.~\ref{fig:edge_dynamics},
\begin{equation}
\begin{aligned}
	&P_1=(-1.60,0.50\pi),\quad
	&P_2=(-2.50,0.125\pi), \\
	&P_3=(-0.10,0.1875\pi),\quad
	&P_4=(-1.65,0.0625\pi),
	\label{eq:P1234}
\end{aligned}
\end{equation}
which correspond respectively to $(\nu_0,\nu_\pi)=(0,0)$, $(0,1)$, $(1,0)$, and $(1,1)$. The four representative points used for the frame-resolved bulk probes in Fig.~\ref{fig:mcd_representative} are
\begin{equation}
\begin{aligned}
	&Q_1=(-1.50,\pi/3),\quad
	&Q_2=(0,\pi/3),\quad\\
	&Q_3=(-2.50,0),\quad
	&Q_4=(-1.50,0),
	\label{eq:Q1234}
\end{aligned}
\end{equation}
which realize $(W_1,W_2)=(0,0)$, $(1,1)$, $(1,-1)$, and $(2,0)$. The two benchmark critical points used in Fig.~\ref{fig:critical_benchmark} are
\begin{equation}
C_0=(0,\pi/2),
\qquad
C_\pi=(\pi-2,0),
\label{eq:C0Cpi}
\end{equation}
where $C_0$ is a representative $0$ gap closing and $C_\pi$ is a representative $\pi$ gap closing. Labeling these points explicitly removes the ambiguity between the phase diagram and the later dynamical panels.

\section{Open-boundary edge sector and the $0/\pi$ logical subspace}

Under open boundary conditions, the coexistence sector supports boundary states pinned near quasienergies $0$ and $\pi/T$. On the left edge we denote these two anomalous modes by $\ket{L,0}$ and $\ket{L,\pi}$,
with Floquet eigenvalue equations
\begin{equation}
\UT\ket{L,0}=+\ket{L,0},
\qquad
\UT\ket{L,\pi}=-\ket{L,\pi},
\label{eq:edge_pair_eigs}
\end{equation}
up to exponentially small finite size corrections.

\begin{figure*}[t]
\centering
\includegraphics[width=0.8\textwidth]{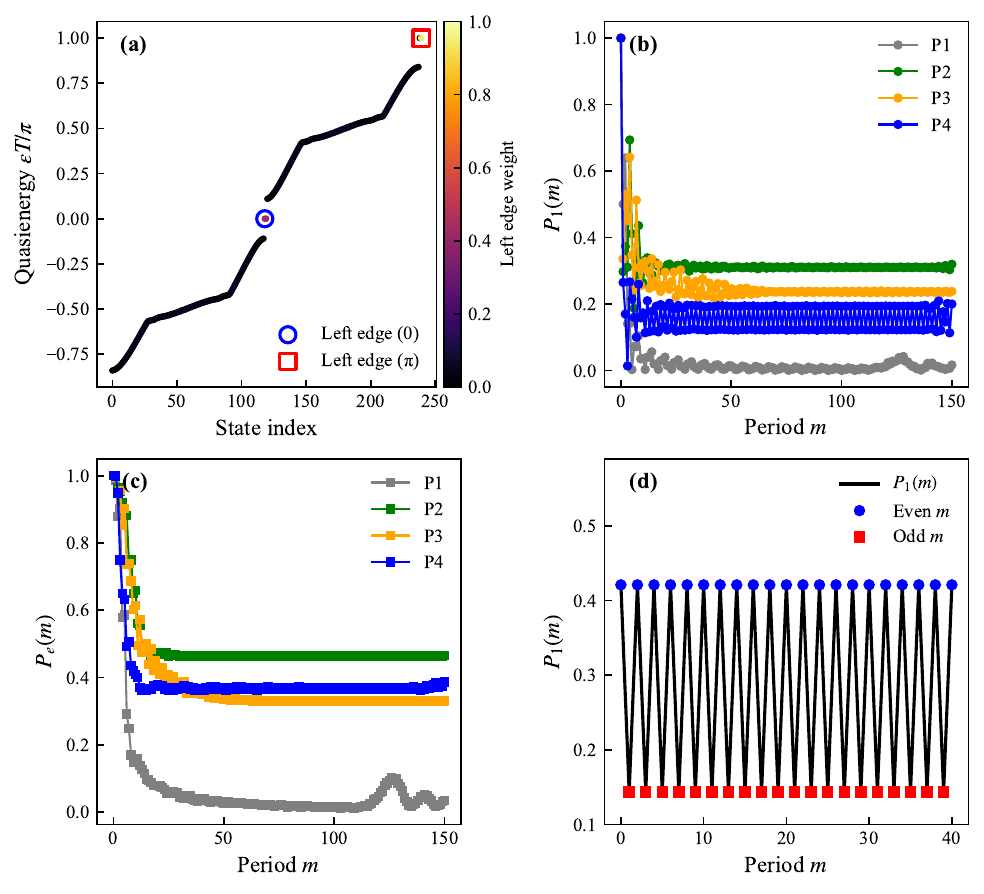}
\caption{Open-boundary edge spectrum and boundary dynamics. Panel (a) shows the quasienergy spectrum at $P_4$ under open boundary conditions. The color scale gives the left edge weight in a fixed boundary window, and the highlighted markers identify the left edge states near $0$ and $\pi/T$ on the same boundary. Panel (b) shows the site-one probability $P_1(m)$ defined in Eq.~\eqref{eq:P1Pe_defs} for the four points $P_1$ to $P_4$ defined in Fig.~\ref{fig:bulk_phase_map}, with the initial state $\ket{\Psi(0)}=\ket{1,A}$. Panel (c) shows the boundary window probability $P_{\mathrm e}(m)$ defined in Eq.~\eqref{eq:P1Pe_defs} for the same four points. Panel (d) shows the optimized doubled-period response at $P_4$ for the phase matched state in Eq.~\eqref{eq:optimized_edge_superposition}. The continuous trace is $P_1(m)$ itself, while the circular and square markers identify the even and odd periods. The explicit step-by-step alternation is the direct numerical realization of Eq.~\eqref{eq:period_doubled_response}.}
\label{fig:edge_dynamics}
\end{figure*}

For the later numerical panels we define two explicit boundary observables. The site-one projector is
\begin{equation}
\Pi_1=\ket{1,A}\!\bra{1,A}+\ket{1,B}\!\bra{1,B},
\label{eq:Pi1_def}
\end{equation}
and the boundary window projector is
\begin{equation}
\Pi_{\mathrm e}=\sum_{n=1}^{n_{\mathrm e}}\qty(\ket{n,A}\!\bra{n,A}+\ket{n,B}\!\bra{n,B}),
\label{eq:Pie_def}
\end{equation}
where $n_{\mathrm e}$ is the number of unit cells retained in the edge window. For any stroboscopic state $\ket{\Psi(m)}=\UT^m\ket{\Psi(0)}$ we then define
\begin{equation}
	\begin{split}
		&P_1(m)=\mel{\Psi(m)}{\Pi_1}{\Psi(m)},\\
		&P_{\mathrm e}(m)=\mel{\Psi(m)}{\Pi_{\mathrm e}}{\Psi(m)}.
		\label{eq:P1Pe_defs}
	\end{split}
\end{equation}
In panels (b) and (c) of Fig.~\ref{fig:edge_dynamics} the initial state is chosen as the left boundary cell state $\ket{\Psi(0)}=\ket{1,A}$.

The two anomalous edge modes span a natural boundary logical subspace,
\begin{equation}
\mathcal H_{\mathrm{edge}}^{(L)}=\mathrm{span}\{\ket{L,0},\ket{L,\pi}\},
\label{eq:Hedge_def}
\end{equation}
with the corresponding logical basis
\begin{equation}
\ket{0}_{\mathrm L}=\ket{L,0},
\qquad
\ket{1}_{\mathrm L}=\ket{L,\pi}.
\label{eq:logical_basis_def}
\end{equation}
This boundary subspace is not an accidental two-level sector. Its existence is tied to the chiral symmetry of the Floquet walk, with $\Gamma=\sigma_y$ in Eq.~\eqref{eq:Gamma_def}, and to the simultaneous opening of the quasienergy gaps at $0$ and $\pi/T$. Local perturbations that preserve the chiral symmetry and do not close either gap cannot remove the two edge modes individually; they can only deform their spatial profiles and modify exponentially small finite-size splittings. In this restricted but physically important sense, $\mathcal H_{\mathrm{edge}}^{(L)}$ is a symmetry-protected boundary logical sector. The term edge qubit used below refers to this protected Floquet subspace, rather than to a claim of universal fault-tolerant topological quantum computation.

In this basis,
\begin{equation}
\UT\big|_{\mathcal H_{\mathrm{edge}}^{(L)}}=
\begin{pmatrix}
1&0\\
0&-1
\end{pmatrix}
\equiv \tau_z.
\label{eq:edge_Z_gate}
\end{equation}
After $m$ periods,
\begin{equation}
\UT^m\big|_{\mathcal H_{\mathrm{edge}}^{(L)}}=\tau_z^m.
\label{eq:edge_Z_gate_m}
\end{equation}
Once the Floquet action has been reduced to Eqs.~\eqref{eq:edge_Z_gate} and \eqref{eq:edge_Z_gate_m}, it is natural to regard this two-level logical subspace as an edge qubit operated stroboscopically by the anomalous $0/\pi$ phase structure. Therefore a coherent superposition,
\begin{equation}
\ket{\psi_{\mathrm{edge}}(0)}=\alpha\ket{L,0}+\beta\ket{L,\pi},
\label{eq:edge_superposition}
\end{equation}
evolves into
\begin{equation}
\ket{\psi_{\mathrm{edge}}(m)}=\alpha\ket{L,0}+(-1)^m\beta\ket{L,\pi}.
\label{eq:edge_evolution_superposition}
\end{equation}
The relative sign alternates from one period to the next. This is the microscopic origin of the doubled-period response.

For a local boundary observable $\hat O_{\mathrm{edge}}$,
\begin{align}
\expval{\hat O_{\mathrm{edge}}}_m={}&|\alpha|^2\matrixel{L,0}{\hat O_{\mathrm{edge}}}{L,0}
+|\beta|^2\matrixel{L,\pi}{\hat O_{\mathrm{edge}}}{L,\pi}\nonumber\\
&+(-1)^m\alpha^*\beta\matrixel{L,0}{\hat O_{\mathrm{edge}}}{L,\pi}\nonumber\\
&+(-1)^m\beta^*\alpha\matrixel{L,\pi}{\hat O_{\mathrm{edge}}}{L,0}.
\label{eq:Oedge_expanded}
\end{align}
Grouping the constant and alternating pieces gives
\begin{equation}
\expval{\hat O_{\mathrm{edge}}}_m=O_{\mathrm{dc}}+(-1)^m O_{\mathrm{ac}}.
\label{eq:period_doubled_response}
\end{equation}
Equation \eqref{eq:period_doubled_response} already shows that a robust $2T$ signal requires simultaneous overlap with the $0$ and $\pi$ sectors. This point becomes sharper if one decomposes a generic boundary localized initial state as
\begin{equation}
\ket{\Psi(0)}=c_0\ket{L,0}+c_\pi\ket{L,\pi}+\sum_{\mu} c_\mu \ket{\mu},
\label{eq:generic_edge_decomp}
\end{equation}
where $\ket{\mu}$ denotes every remaining Floquet eigenstate with
\begin{equation}
\UT\ket{\mu}=\e^{-\ii\varepsilon_\mu T}\ket{\mu}.
\label{eq:bulk_modes_def}
\end{equation}
For any boundary local observable $\hat O_{\mathrm{edge}}$ one then has
\begin{align}
\expval{\hat O_{\mathrm{edge}}}_m
=
&|c_0|^2 O_{00}
+|c_\pi|^2 O_{\pi\pi}\nonumber\\
&+(-1)^m\qty(c_0^*c_\pi O_{0\pi}+\mathrm{c.c.})
+R(m),
\label{eq:edge_general_response}
\end{align}
where $R(m)$ collects all terms involving at least one nonanomalous mode. Once the fast oscillatory part $R(m)$ dephases, the long-time boundary response falls into four distinct classes,
\begin{equation}
\expval{\hat O_{\mathrm{edge}}}_m \longrightarrow
\begin{cases}
0\ \text{or a small background}, & \text{trivial sector},\\[3pt]
|c_0|^2 O_{00}, & 0\text{-only sector},\\[3pt]
|c_\pi|^2 O_{\pi\pi}, & \pi\text{-only sector},\\[3pt]
O_{\mathrm{dc}}+(-1)^m O_{\mathrm{ac}}, & 0+\pi\text{ sector}.
\end{cases}
\label{eq:sector_dependent_periods}
\end{equation}
The crucial point is that a single $0$ edge mode or a single $\pi$ edge mode produces a stroboscopically constant local expectation value, because the overall Floquet phase cancels in $\expval{\hat O_{\mathrm{edge}}}_m$. A genuine $2T$ signal appears only when both anomalous edge sectors are coherently occupied on the same boundary. Appendix~\ref{app:edge_periods} gives the detailed derivation.

Figure \ref{fig:edge_dynamics} gives the numerical realization of this structure. Panel (a) shows the open boundary quasienergy spectrum at $P_4$ and identifies the left edge states near $0$ and $\pi/T$. Panels (b) and (c) compare $P_1(m)$ and $P_{\mathrm e}(m)$ for the four labeled points $P_1$ to $P_4$. The trivial point loses its local boundary weight. The $0$-only and $\pi$-only points settle into stroboscopically constant boundary signals with different plateau values. The coexistence point retains both anomalous sectors and therefore supports a doubled-period response. In panel (d) we choose the phase matched superposition
\begin{equation}
\ket{\psi_{\mathrm opt}(0)}=
\frac{1}{\sqrt2}\qty(\ket{L,0}+\e^{-\ii\arg\matrixel{L,0}{\Pi_1}{L,\pi}}\ket{L,\pi}),
\label{eq:optimized_edge_superposition}
\end{equation}
which maximizes the alternating contribution of $\Pi_1$. The resulting short time trace of $P_1(m)$ shows the $2T$ response directly as a period-by-period alternation.

The coexistence sector therefore provides more than an enlarged boundary spectrum. It provides a genuine two-dimensional boundary subspace with a definite Floquet phase action, and it is only this coexistence structure that gives a robust doubled-period signal in local boundary observables.

\section{Bulk dynamical probes of anomalous Floquet topology}

The bulk part of the walk should be read in the same language as the edge part. The main question is not whether one can draw another propagation plot, but whether the anomalous Floquet topology leaves direct dynamical fingerprints in bulk observables. In the present model this information appears in two complementary observables: a frame-resolved topological probe and a benchmark comparison of local stroboscopic dynamics at representative $0$ gap and $\pi$ gap critical points.

\subsection{Frame-resolved mean chiral displacement}

\begin{figure*}[t]
\centering
\includegraphics[width=\textwidth]{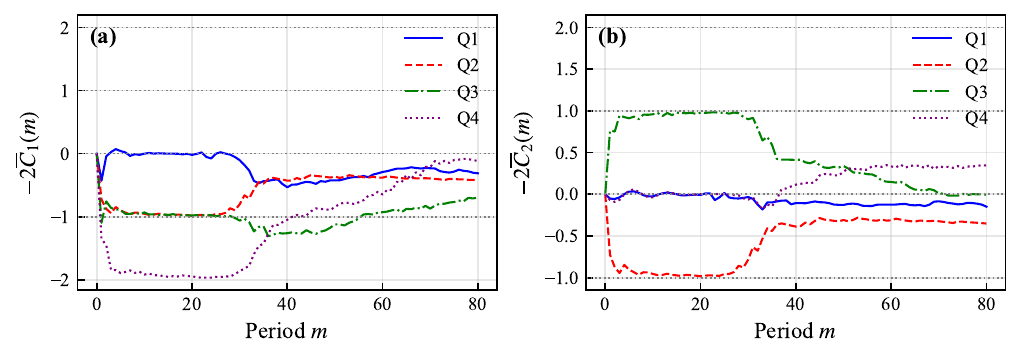}
\caption{Frame-resolved mean chiral displacement for the four representative bulk points $Q_1$ to $Q_4$ defined in Eq.~\eqref{eq:Q1234}. Panel (a) shows $-2\Cbar_1(m)$ in symmetric time frame 1 and panel (b) shows $-2\Cbar_2(m)$ in symmetric time frame 2. The dotted horizontal guides mark the quantized targets associated with the winding numbers. Over the pre-reflection time window the curves approach $(W_1,W_2)=(0,0)$ for the trivial point, $(1,1)$ for the $0$-only point, $(1,-1)$ for the $\pi$-only point, and $(2,0)$ for the coexistence point. The two-gap bulk topology is therefore dynamically resolved in the two symmetric time frames.}
\label{fig:mcd_representative}
\end{figure*}

Because the walk is chiral in two symmetric time frames, a natural bulk probe is the frame-resolved mean chiral displacement. For the numerical data in Fig.~\ref{fig:mcd_representative} we work on a long periodic chain of odd length $L=2L_0+1$, choose the central launch cell
\begin{equation}
n_0=\frac{L+1}{2},
\label{eq:n0_center}
\end{equation}
and take the initial bulk localized state
\begin{equation}
\ket{\psi_0}=\ket{n_0,A}.
\label{eq:psi0_mcd}
\end{equation}
The position operator is measured relative to this preparation center,
\begin{equation}
\hat x=\sum_{n=1}^{L}(n-n_0)\qty(\ket{n,A}\!\bra{n,A}+\ket{n,B}\!\bra{n,B}),
\label{eq:x_operator_def}
\end{equation}
and $\Gamma=\sigma_y$ acts in the coin subspace. Let $U_\ell$ denote the Floquet operator in symmetric time frame $\ell=1,2$. We define
\begin{equation}
C_\ell(m)=\mel{\psi_0}{(U_\ell^\dagger)^m\,\hat x\,\Gamma\,U_\ell^m}{\psi_0}
\qquad \ell=1,2.
\label{eq:Cl_def}
\end{equation}
To suppress finite time oscillations, we use the running average
\begin{equation}
\Cbar_\ell(m)=\frac{1}{m+1}\sum_{r=0}^{m} C_\ell(r).
\label{eq:running_average_def}
\end{equation}
In our convention the long-time limits satisfy
\begin{equation}
-2\Cbar_1(m)\to W_1,
\quad
-2\Cbar_2(m)\to W_2,
\quad m\to\infty.
\label{eq:MCD_quantization}
\end{equation}
This relation is the dynamical bulk counterpart of Eqs.~\eqref{eq:winding_def} and \eqref{eq:nu_def}. A step-by-step momentum space derivation of the time-averaged part is summarized in Appendix~\ref{app:mcd_appendix}.

The asymptotic relation in Eq.~\eqref{eq:MCD_quantization} is the thermodynamic target, while Fig.~\ref{fig:mcd_representative} shows how the finite time traces approach those targets over the clean pre-reflection window for the four representative points $Q_1$ to $Q_4$. The trivial point $Q_1$ approaches $(0,0)$. The $0$-only point $Q_2$ approaches $(1,1)$. The $\pi$-only point $Q_3$ approaches $(1,-1)$. The coexistence point $Q_4$ approaches $(2,0)$. The agreement is direct and does not rely on open boundaries. The dotted horizontal guides mark the corresponding winding targets. This provides bulk dynamical evidence that the system is an anomalous Floquet quantum walk with two independent winding numbers.

\subsection{Selected benchmark critical dynamics}

The benchmark bulk probe addresses a different question. Even though both $0$ gap and $\pi$ gap closings are encoded in the same phase diagram, they need not produce identical local stroboscopic dynamics. To make this point cleanly, we do not claim a uniform behavior along every critical branch. Instead we focus on two analytically transparent benchmark cuts, one for a representative $0$ gap closing and one for a representative $\pi$ gap closing.

\begin{figure*}[t]
\centering
\includegraphics[width=\textwidth]{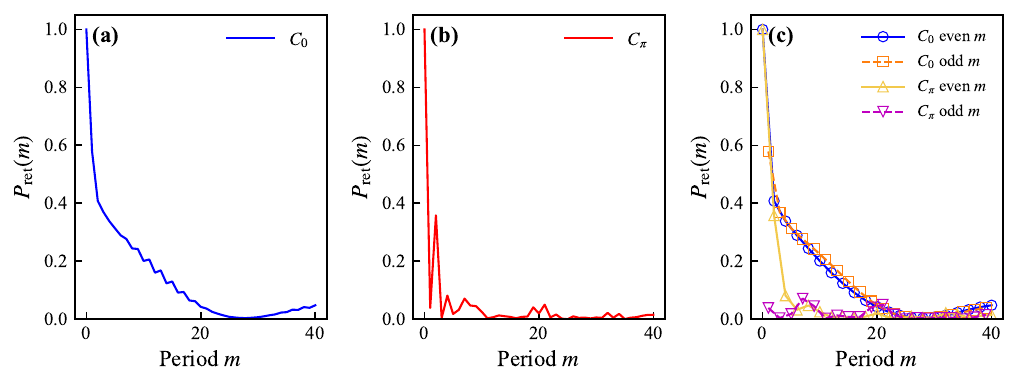}
\caption{Selected benchmark bulk critical dynamics for the two points defined in Eq.~\eqref{eq:C0Cpi}. Panels (a) and (b) show the return probability $P_{\mathrm{ret}}(m)$ defined in Eq.~\eqref{eq:Pret_def} for the representative $0$ gap benchmark $C_0=(0,\pi/2)$ and the representative $\pi$ gap benchmark $C_\pi=(\pi-2,0)$. Panel (c) separates the even and odd subsequences for both benchmarks over the same short time window and compares the local stroboscopic fingerprints directly. The odd and even splitting is much stronger for the $\pi$ gap benchmark, consistent with the leading expansion in Eq.~\eqref{eq:U_Cpi_expansion}. The benchmark figure is therefore focused entirely on the local stroboscopic evidence for the distinction between the two representative critical cuts.}
\label{fig:critical_benchmark}
\end{figure*}

For the benchmark calculation we again use a long periodic chain and the bulk localized launch state
\begin{equation}
\ket{\Psi(0)}=\ket{n_0,A},
\label{eq:psi0_critical}
\end{equation}
with the same central cell $n_0$ as in Eq.~\eqref{eq:n0_center}. The local return projector is
\begin{equation}
\Pi_{n_0}=\ket{n_0,A}\!\bra{n_0,A}+\ket{n_0,B}\!\bra{n_0,B},
\label{eq:Pi_return_def}
\end{equation}
and the return probability used in Fig.~\ref{fig:critical_benchmark} is
\begin{equation}
	\begin{split}
		&P_{\mathrm{ret}}(m)=\mel{\Psi(m)}{\Pi_{n_0}}{\Psi(m)},\\
		&\ket{\Psi(m)}=\UT^m\ket{\Psi(0)}.
		\label{eq:Pret_def}
	\end{split}
\end{equation}

For the $0$ gap benchmark $C_0=(0,\pi/2)$, the gap closes at $k_*=\pi/2$. Writing $k=k_*+q$ with small $q$, one finds
\begin{equation}
\theta_1\simeq 2q,
\qquad
\theta_2\simeq -2q,
\label{eq:theta_C0_expansion}
\end{equation}
so that the Floquet operator expands as
\begin{equation}
U(k_*+q)\simeq \I-\ii\,2q\qty(\sigma_z-\sigma_x).
\label{eq:U_C0_expansion}
\end{equation}
The critical contribution is therefore organized around $+\I$. No intrinsic factor $(-1)^m$ is generated by the stroboscopic evolution.

For the $\pi$ gap benchmark $C_\pi=(\pi-2,0)$, the gap closes at $k_*=0$. Expanding around $k=q$ gives
\begin{equation}
\theta_1\simeq 2q,
\qquad
\theta_2\simeq \pi-q^2,
\label{eq:theta_Cpi_expansion}
\end{equation}
which leads to
\begin{equation}
U(q)\simeq -\I+\ii\,2q\sigma_z+\mathcal O(q^2).
\label{eq:U_Cpi_expansion}
\end{equation}
The key difference is the overall factor $-\I$. It implies that the critical contribution naturally carries a stroboscopic factor $(-1)^m$, so odd and even periods are expected to separate much more clearly in local observables. The Taylor expansions used in Eqs.~\eqref{eq:U_C0_expansion} and \eqref{eq:U_Cpi_expansion} are derived explicitly in Appendix~\ref{app:mcd_appendix}.

Figure \ref{fig:critical_benchmark} shows the numerical comparison. Panels (a) and (b) plot $P_{\mathrm{ret}}(m)$ for the two benchmark points. Panel (c) resolves the even and odd subsequences explicitly. The $\pi$ gap benchmark shows a much stronger odd and even separation, in agreement with Eq.~\eqref{eq:U_Cpi_expansion}, while the $0$ gap benchmark remains much less staggered. We therefore focus the benchmark figure entirely on the local stroboscopic evidence and do not mix it with more global spreading diagnostics.

Taken together, the two bulk probes play different roles. The frame-resolved mean chiral displacement provides a bulk dynamical route toward the two winding numbers, while the benchmark critical dynamics shows that selected $0$ gap and $\pi$ gap cuts are also distinguished in local stroboscopic evolution. These results give the bulk part of the paper the same level of physical precision as the boundary $0/\pi$ logical subspace discussed above.

\section{Conclusion}

We have formulated a flux-controlled anomalous Floquet quantum walk and connected it to a concrete driven bipartite lattice, so that the same step operator controls the dual gap topology, the boundary $0/\pi$ sector, and the bulk dynamical probes. In the coexistence region, the $0$ and $\pi$ edge modes on the same boundary span a symmetry-protected logical subspace whose relative Floquet phase produces the doubled-period local response. In the bulk, frame-resolved mean chiral displacements approach the winding targets in the two symmetric time frames over the clean pre-reflection window, while selected $0$ gap and $\pi$ gap benchmarks exhibit distinct local stroboscopic fingerprints with a stronger odd-even alternation at the $\pi$ gap benchmark. The required ingredients are experimentally concrete: one-dimensional driven bipartite lattices with controllable hopping phases can be approached in periodically driven optical lattices, photonic quantum-walk architectures, and coupled waveguide or resonator arrays, where site-resolved density, boundary return probability, and mean chiral displacement are natural observables. The model therefore goes beyond a generic demonstration of boundary modes or packet propagation: it gives a minimal microscopic anomalous Floquet quantum walk in which boundary logical structure and bulk dynamical readout are tied to the same flux-controlled protocol.

\acknowledgments 
We acknowledge the support of the National Natural Science Foundation of China (Grants No. 12275193, and 11975166).

\appendix

\section{Detailed derivation of the Floquet quantum walk mapping}
\label{app:walk_mapping}

Start from the real space Hamiltonians in Eqs.~\eqref{eq:H1real} and \eqref{eq:H2real}. Introduce the Fourier transforms
\begin{equation}
a_n=\frac{1}{\sqrt L}\sum_k \e^{\ii kn}a_k,
\qquad
b_n=\frac{1}{\sqrt L}\sum_k \e^{\ii kn}b_k.
\label{eq:FT_appendix}
\end{equation}
Then
\begin{align}
\sum_n a_{n+1}^\dagger a_n
={}&\frac{1}{L}\sum_{n,k,q}\e^{-\ii q(n+1)}\e^{\ii kn}a_q^\dagger a_k
=\sum_k \e^{-\ii k}a_k^\dagger a_k,
\label{eq:fourier_a_shift}
\end{align}
and similarly
\begin{equation}
\sum_n a_n^\dagger a_{n+1}=\sum_k \e^{\ii k}a_k^\dagger a_k.
\label{eq:fourier_a_shift_hc}
\end{equation}
Therefore the $a$ part of $H_1$ is
\begin{align}
-\ii t_1\sum_k\qty(\e^{-\ii\phi}\e^{-\ii k}-\e^{\ii\phi}\e^{\ii k})a_k^\dagger a_k
={}&2t_1\sum_k\sin(k-\phi)a_k^\dagger a_k.
\label{eq:H1_a_k}
\end{align}
The $b$ part contributes the opposite sign,
\begin{equation}
-2t_1\sum_k\sin(k-\phi)b_k^\dagger b_k.
\label{eq:H1_b_k}
\end{equation}
Hence
\begin{equation}
H_1(k)=2t_1\sin(k-\phi)\sigma_z.
\label{eq:H1k_appendix}
\end{equation}
For $H_2$, one finds
\begin{equation}
\sum_n a_n^\dagger b_n=\sum_k a_k^\dagger b_k,
\label{eq:onsite_mix_appendix}
\end{equation}
and
\begin{equation}
	\begin{split}
		&\sum_n a_{n+1}^\dagger b_n=\sum_k \e^{-\ii k}a_k^\dagger b_k,
		\\
		&\sum_n a_n^\dagger b_{n+1}=\sum_k \e^{\ii k}a_k^\dagger b_k.
		\label{eq:offdiag_mix_appendix}
	\end{split}
\end{equation}
Therefore
\begin{align}
	H_2(k)&=\qty(M+t_2\e^{-\ii k}+t_2\e^{\ii k})\sigma_x
	\nonumber
	\\
	&=\qty(M+2t_2\cos k)\sigma_x.
	\label{eq:H2k_appendix}
\end{align}
Inserting Eqs.~\eqref{eq:H1k_appendix} and \eqref{eq:H2k_appendix} into Eq.~\eqref{eq:UFdef} gives Eq.~\eqref{eq:Ukcompact}. These two contributions generate, respectively, the coin-dependent drift and the coin mixing process.

\section{Detailed derivation of the quasienergy bands and gap closings}
\label{app:bands_gaps}

Start from
\begin{equation}
U(k)=\e^{-\ii\theta_2\sigma_x}\e^{-\ii\theta_1\sigma_z}.
\label{eq:U_start_appendix}
\end{equation}
Use
\begin{equation}
	\begin{split}
		\e^{-\ii\theta_2\sigma_x}=\cos\theta_2\,\I-\ii\sin\theta_2\,\sigma_x,
		\\
		\e^{-\ii\theta_1\sigma_z}=\cos\theta_1\,\I-\ii\sin\theta_1\,\sigma_z.
		\label{eq:exp_pauli_appendix}
	\end{split}
\end{equation}
Multiplying the two factors yields
\begin{align}
U(k)={}&\cos\theta_2\cos\theta_1\,\I
-\ii\cos\theta_2\sin\theta_1\,\sigma_z\nonumber
\\&-\ii\sin\theta_2\cos\theta_1\,\sigma_x 
-\sin\theta_2\sin\theta_1\,\sigma_x\sigma_z.
\label{eq:U_mult_appendix}
\end{align}
Because
\begin{equation}
\sigma_x\sigma_z=-\ii\sigma_y,
\label{eq:sigmaxsigmaz_appendix}
\end{equation}
Eq.~\eqref{eq:Uexpanded} follows immediately. Taking the trace gives
\begin{equation}
\Tr U(k)=2\cos\theta_2\cos\theta_1.
\label{eq:traceU_appendix}
\end{equation}
If the eigenvalues are $\e^{-\ii\varepsilon T}$ and $\e^{\ii\varepsilon T}$, then
\begin{equation}
\Tr U(k)=2\cos(\varepsilon T).
\label{eq:trace_from_eigs_appendix}
\end{equation}
Equating Eqs.~\eqref{eq:traceU_appendix} and \eqref{eq:trace_from_eigs_appendix} gives Eq.~\eqref{eq:cosdisp}. The closing conditions follow by imposing $\cos(\varepsilon T)=\pm1$, which means $\varepsilon T=\ell\pi$. Since $\cos\theta_1\cos\theta_2=\pm1$ and each factor has magnitude at most one, each factor must separately be $\pm1$, yielding Eq.~\eqref{eq:gapclosing_explicit}.

\section{Detailed derivation of the winding numbers in the two symmetric time frames}
\label{app:winding_derivation}

Consider the frame
\begin{equation}
U_1=\e^{-\ii\frac{\theta_1}{2}\sigma_z}\e^{-\ii\theta_2\sigma_x}\e^{-\ii\frac{\theta_1}{2}\sigma_z}.
\label{eq:U1_appendix_start}
\end{equation}
Multiplying the left pair of factors gives
\begin{align}
	& \e^{-\ii\frac{\theta_1}{2}\sigma_z}\e^{-\ii\theta_2\sigma_x} \nonumber\\
	={}&\qty(\cos\frac{\theta_1}{2}\I-\ii\sin\frac{\theta_1}{2}\sigma_z)
	\qty(\cos\theta_2\I-\ii\sin\theta_2\sigma_x) \nonumber\\
	={}&\cos\frac{\theta_1}{2}\cos\theta_2\I
	-\ii\cos\frac{\theta_1}{2}\sin\theta_2\sigma_x\nonumber\\
	&-\ii\sin\frac{\theta_1}{2}\cos\theta_2\sigma_z 
	+\ii\sin\frac{\theta_1}{2}\sin\theta_2\sigma_y.
	\label{eq:U1_first_product}
\end{align}
Multiplying by the final factor and simplifying yields
\begin{equation}
U_1=n_{0,1}\I-\ii\qty(n_{x,1}\sigma_x+n_{z,1}\sigma_z),
\label{eq:U1_final_appendix}
\end{equation}
with the coefficients quoted in Eqs.~\eqref{eq:n1x} and \eqref{eq:n1z}. The same procedure for $U_2$ gives Eqs.~\eqref{eq:n2x} and \eqref{eq:n2z}. Since the chiral operator is $\Gamma=\sigma_y$, the relevant vectors lie in the $x$-$z$ plane. As $k$ runs through the Brillouin zone, the complex number $n_x(k)+\ii n_z(k)$ winds around the origin an integer number of times. That integer is Eq.~\eqref{eq:winding_def}. The combinations in Eq.~\eqref{eq:nu_def} then follow from the standard bulk edge correspondence for one-dimensional chiral Floquet walks~\cite{asbothBulkBoundaryCorrespondenceChiral2013}.

\section{Detailed derivation of the boundary phase operation and the sector dependent local periods}
\label{app:edge_periods}

The boundary discussion in the main text contains two logically distinct parts. One part shows why a coherent $0/\pi$ superposition generates a doubled-period signal. The other part shows why the four topological sectors produce different long-time boundary responses in local observables.

Because
\begin{equation}
\UT\ket{L,0}=+\ket{L,0},
\qquad
\UT\ket{L,\pi}=-\ket{L,\pi},
\label{eq:D_edgepair}
\end{equation}
a superposition of the two anomalous edge modes,
\begin{equation}
\ket{\psi(0)}=\alpha\ket{L,0}+\beta\ket{L,\pi},
\label{eq:D_state}
\end{equation}
evolves into
\begin{equation}
\ket{\psi(m)}=\alpha\ket{L,0}+(-1)^m\beta\ket{L,\pi}.
\label{eq:D_state_evolved}
\end{equation}
For a local boundary observable $\hat O_{\mathrm{edge}}$ one then obtains
\begin{align}
\expval{\hat O_{\mathrm{edge}}}_m
={}&\mel{\psi(m)}{\hat O_{\mathrm{edge}}}{\psi(m)} \nonumber\\
={}&|\alpha|^2 O_{00}+|\beta|^2 O_{\pi\pi}\nonumber\\
&+(-1)^m\alpha^*\beta O_{0\pi}
+(-1)^m\beta^*\alpha O_{\pi 0},
\label{eq:D_O_expanded}
\end{align}
where
\begin{equation}
	\begin{split}
		&O_{00}=\matrixel{L,0}{\hat O_{\mathrm{edge}}}{L,0},
		\\
		&O_{\pi\pi}=\matrixel{L,\pi}{\hat O_{\mathrm{edge}}}{L,\pi},
		\\
		&O_{0\pi}=\matrixel{L,0}{\hat O_{\mathrm{edge}}}{L,\pi}.
		\label{eq:O_matrix_elements_appendix}
	\end{split}
\end{equation}
Therefore
\begin{equation}
\expval{\hat O_{\mathrm{edge}}}_m=O_{\mathrm{dc}}+(-1)^m O_{\mathrm{ac}},
\label{eq:dc_ac_appendix}
\end{equation}
which is the origin of the doubled-period response.

To classify the long-time local responses of the different topological sectors, we must now include all other Floquet eigenstates. Let
\begin{equation}
\ket{\Psi(0)}=c_0\ket{L,0}+c_\pi\ket{L,\pi}+\sum_\mu c_\mu \ket{\mu},
\label{eq:D_full_decomp}
\end{equation}
where the remaining eigenstates satisfy
\begin{equation}
\UT\ket{\mu}=\e^{-\ii\varepsilon_\mu T}\ket{\mu}.
\label{eq:D_bulk_eigs}
\end{equation}
The evolved state is
\begin{align}
	\ket{\Psi(m)}=&c_0\ket{L,0}+(-1)^m c_\pi\ket{L,\pi}\nonumber\\
	&+\sum_\mu c_\mu \e^{-\ii m\varepsilon_\mu T}\ket{\mu}.
	\label{eq:D_full_evolved}
\end{align}
Insert Eq.~\eqref{eq:D_full_evolved} into $\expval{\hat O_{\mathrm{edge}}}_m$. The terms containing only the two anomalous edge modes reproduce Eq.~\eqref{eq:dc_ac_appendix}. Every remaining contribution contains at least one nonanomalous mode and can be grouped into a remainder

\begin{align}
	R(m)=
	&\sum_\mu \Bigl[c_0^*c_\mu \e^{-\ii m\varepsilon_\mu T} O_{0\mu}
	+c_\pi^*c_\mu (-1)^m \e^{-\ii m\varepsilon_\mu T} O_{\pi\mu}\nonumber\\
	&+\mathrm{c.c.}\Bigr]
	+\sum_{\mu,\nu} c_\mu^* c_\nu \e^{\ii m(\varepsilon_\mu-\varepsilon_\nu)T} O_{\mu\nu},
	\label{eq:D_remainder}
\end{align}

with the obvious notation $O_{ab}=\matrixel{a}{\hat O_{\mathrm{edge}}}{b}$. Hence
\begin{align}
\expval{\hat O_{\mathrm{edge}}}_m=
&|c_0|^2 O_{00}
+|c_\pi|^2 O_{\pi\pi}\nonumber\\
&+(-1)^m\qty(c_0^*c_\pi O_{0\pi}+\mathrm{c.c.})
+R(m).
\label{eq:D_master_local}
\end{align}

Equation~\eqref{eq:D_master_local} immediately explains the sector dependence of local boundary observables. In the trivial sector there is no anomalous edge mode, so the boundary localized initial state overlaps only nonanomalous states and the long-time local edge signal is lost once the packet leaks into the bulk. In the $0$-only sector one has $c_\pi=0$, and the only persistent edge contribution is $|c_0|^2 O_{00}$, which is stroboscopically constant. In the $\pi$-only sector one has $c_0=0$, and the only persistent edge contribution is $|c_\pi|^2 O_{\pi\pi}$, which is also stroboscopically constant because the single Floquet phase $\e^{-\ii \pi m}$ cancels out in the expectation value. Only in the coexistence sector can both $c_0$ and $c_\pi$ be nonzero on the same edge, and only then does the alternating term survive. The long-time classification quoted in Eq.~\eqref{eq:sector_dependent_periods} follows directly.

The optimized state used in Fig.~\ref{fig:edge_dynamics}(d) follows from the same formula. For $\hat O_{\mathrm{edge}}=\Pi_1$, the alternating amplitude is
\begin{equation}
O_{\mathrm{ac}}=\alpha^*\beta \matrixel{L,0}{\Pi_1}{L,\pi}+\mathrm{c.c.}
\label{eq:D_opt_amp}
\end{equation}
under the constraint $|\alpha|=|\beta|=1/\sqrt2$. This amplitude is maximal when
\begin{equation}
\beta=\frac{1}{\sqrt2}\e^{-\ii\arg\matrixel{L,0}{\Pi_1}{L,\pi}},
\qquad
\alpha=\frac{1}{\sqrt2},
\label{eq:D_phase_match}
\end{equation}
which reproduces Eq.~\eqref{eq:optimized_edge_superposition}.

\section{Bulk dynamical probes in the symmetric time frames and at benchmark critical points}
\label{app:mcd_appendix}

For the frame-resolved bulk probe, translation invariance allows a momentum space derivation of the time-averaged part. For a centered local initial state, the momentum amplitude is uniform, and the position operator acts as $\hat x=\ii\partial_k$ in Bloch space. Therefore
\begin{equation}
C_\ell(m)=\int_{-\pi}^{\pi}\frac{dk}{2\pi}\,
\mel{\chi_0}{U_\ell^{-m}(k)\,\ii\partial_k\!\qty[\Gamma U_\ell^m(k)]}{\chi_0},
\label{eq:E_mcd_kspace}
\end{equation}
where $\ket{\chi_0}$ is the coin part of the initial state. In the symmetric time frames the Bloch operator can be written as
\begin{equation}
	\begin{split}
		&U_\ell(k)=\cos E_\ell(k)\,\I-\ii\sin E_\ell(k)\,\hat{\bm n}_\ell(k)\cdot\bm\sigma,
		\\
		&\hat{\bm n}_\ell(k)=\frac{\qty(n_{x,\ell}(k),0,n_{z,\ell}(k))}{\sqrt{n_{x,\ell}(k)^2+n_{z,\ell}(k)^2}}.
		\label{eq:E_Uell_axis}
	\end{split}
\end{equation}
The integrand in Eq.~\eqref{eq:E_mcd_kspace} separates into a time independent geometric term and an oscillatory term proportional to $\e^{\pm 2\ii m E_\ell(k)}$. The time independent part is
\begin{align}
C_{\ell,\mathrm{geo}}
&=
-\frac{1}{2}
\int_{-\pi}^{\pi}\frac{dk}{2\pi}\,
\frac{n_{x,\ell}\partial_k n_{z,\ell}-n_{z,\ell}\partial_k n_{x,\ell}}
{n_{x,\ell}^2+n_{z,\ell}^2}\nonumber\\
&=
-\frac{W_\ell}{2},
\label{eq:E_Cgeo}
\end{align}
which fixes the quantized target in our sign convention. The remaining oscillatory term does not vanish at finite time, which is why the raw $C_\ell(m)$ still wiggles in Fig.~\ref{fig:mcd_representative}. The running average in Eq.~\eqref{eq:running_average_def} suppresses exactly this oscillatory contribution and leaves the geometric part in the pre-reflection window. This yields Eq.~\eqref{eq:MCD_quantization}.

The benchmark critical expansions follow directly from Eqs.~\eqref{eq:Ukcompact} and \eqref{eq:theta12}. For the $0$ gap point $C_0=(0,\pi/2)$, set $k=\pi/2+q$ with small $q$. Then
\vspace{-5pt}
\begin{equation}
	\begin{split}
		&\theta_1=Tt_1\sin q\simeq 2q,\\
		&\theta_2=\frac{T}{2}\qty(2\cos\qty(\frac{\pi}{2}+q))\simeq -2q,
		\label{eq:C0_appendix_theta}
	\end{split}
\end{equation}
which yields Eq.~\eqref{eq:U_C0_expansion}. For the $\pi$ gap point $C_\pi=(\pi-2,0)$, set $k=q$. Then
\begin{equation}
	\begin{split}
		&\theta_1=Tt_1\sin q\simeq 2q,
		\\
		&\theta_2=\frac{T}{2}\qty(\pi-2+2\cos q)\simeq \pi-q^2,
		\label{eq:Cpi_appendix_theta}
	\end{split}
\end{equation}
which yields Eq.~\eqref{eq:U_Cpi_expansion}. The different prefactors $\I$ and $-\I$ in the two expansions are the reason why the local stroboscopic response is much more strongly staggered at the selected $\pi$ gap benchmark.

%


%

\end{document}